\documentclass[12pt]{article}
\usepackage{amsmath}
\usepackage{graphicx}

\newcommand{\asi}{A}
\newcommand{\boost}{\Delta}
\newcommand{\hh}{\mathbf{H}}
\newcommand{\ppx}{X}
\newcommand{\prlv}{p}
\newcommand{\qnrlv}{q_N}
\newcommand{\qrlv}{q_R}
\newcommand{\qs}{Q}
\newcommand{\rlv}{R}

\newcommand{\bra}[1]{\pmb{\langle}#1\pmb{|}}

\newcommand{\braket}[2]{
\pmb{\langle}
#1\pmb{\vert}#2\pmb{\rangle}}

\newcommand{\nrlv}{\overline{\rlv}}
\newcommand{\nrlvs}{\bra{\nrlv}}
\newcommand{\qss}{\bra{\qs}}
\newcommand{\rlvs}{\bra{\rlv}}

\title{A toy model of information retrieval system based on quantum probability}
\author{Rom\`an Zapatrin}

\begin{document}

\maketitle

\begin{abstract}
Recent numerical results show that non-Bayesian knowledge revision may be helpful in search engine training and optimization. In order to demonstrate how basic assumption about about the physical nature (and hence the observed statistics) of retrieved documents can affect the performance of search engines we suggest an idealized toy model with minimal number of parameters. 
\end{abstract}


\section*{Introduction}

Two goals are targeted in this opus. The first stems from various indications that IR environments may demonstrate non-classical behavior, similar to that of quantum objects, and we would like to study the emergence of these effects on simple, `tame' models. The second is to operate the idea that the volume of data corpora became so large that it can be treated to be a continuous medium like it is done in solid state physics (this option was highlighted in \cite{me-daniel}) and information retrieval becomes akin to quantum measurement. 

In order to simulate the emergence of quantum behavior we suggest a toy model with few parameters which can be physically based on classical or quantum material. The principal assumption for the model are:
\begin{itemize}
\item the number of retrieved documents is potentially infinite
\item both the relevance (non-relevance) of a document and the occurrence of a particular term $\ppx$ in it are measurable properties, which are subject to physical measurement
\item the properties form an infinite set
\item the method we use to enhance the performance is query expansion: we take a term $\ppx$ and, within a given query, pre-select documents by possessing the term $X$
\end{itemize}

Based on this principles, we suggest a toy model of Information retrieval. More precisely, we consider two models built in a similar way, but with the difference that the first is based on classical objects (macro-objects) governed by Boolean logic, while the second deals with quantum microparticles. 

Then, we carry out a series of numerical experiments with both models, the design of the experiments is similar and we explore the deviation of the internal logic of the second model from the Boolean one. As a measure of this discrepancy we use Accardi statistical invariant \cite{accardi} associated with each term $\ppx$. The results of the numerical evaluations are then compared with the results of a similar experiment performed over TIPSTER test collection \cite{melucci-x}. 

\section{Query expansion}

Let us first exactly describe the settings we assume for our toy model. We are not going to deal with average precision (AP), rather, dwell on a simpler thing: just increasing the precision, saying nothing about the recall. That is, we use strictly one tool: query expansion by pre-filtering of term occurrence, and nothing more. 

\paragraph{Step 1. Initial setting.} First, the relevance is tested. Both relevance and non-relevance are nothing but properties, in our simple model they will correspond to states $\rlvs$, $\nrlvs$, respectively. After the relevance is justified, we check the occurrence of certain term $\ppx$. Again, the term occurrence is nothing but a property, and we check the documents if they possess it.
\unitlength.8mm
\[
\begin{picture}(180,40)
\put(0,20){\framebox(27,27){Source}}
\put(50,20){\framebox(27,27){Relevance}}
\put(100,20){\framebox(27,27){Check $\ppx$}}
\multiput(31,26)(0,5){4}{\vector(1,0){14}}
\multiput(81,26)(0,5){4}{\vector(1,0){14}}
\end{picture}
\]
\paragraph{Step 2. Updated search: query expansion.} Once a term $\ppx$ is set and the appropriate measurements are carried out, the search is modified by \emph{term pre-selection} -- or, in other words, query expansion. This is done as follows: leaving the query, that is, the state $\qss$ the same, we perform a pre-selection and take only those documents which contain the term $\ppx$. 
\[
\begin{picture}(180,40)
\put(0,20){\framebox(27,27){Source}}
\put(50,20){\framebox(27,27){Filter $\ppx$}}
\put(100,20){\framebox(27,27){Relevance}}
\multiput(31,26)(0,5){4}{\vector(1,0){14}}
\multiput(81,26)(0,5){4}{\vector(1,0){14}}
\end{picture}
\]
Then the result of the initial and of the updated queries are compared by forming
\begin{equation}\label{eboost29}
\boost(\ppx)=\frac{P(\rlv\vert\ppx) - P(\rlv)}{P(\rlv)}
\end{equation}
which can take both positive and negative values depending on the choice of the term $\ppx$. 

\section{Melucci Metaphor} 

Melucci metaphor is a unified view to represent simplified IR environment with no reference to particular underlying logic. According to it \cite{melucci2012}, the IR procedure is represented by a two-slit experiment, widely known in physics. The IR system is thought of as a laboratory with the source, which supplies documents according to the input query. The documents within Melucci Metaphor are particles, they may be of classical or quantum nature, or, perhaps, of some other kind. We do not dwell on the mechanism of producing this flux of documents-particles. What is essential, is that the number of ejected documents is supposed to be potentially infinite, but we analyze only first $N$ documents. Putting a particular query means preparing the source in a particular state $\qss$. From this experiment we get the value of $P(\ppx\vert\rlv)$:
\[
\begin{picture}(180,55)
\put(0,30){\oval(40,40)[r]}
\put(21,33){\vector(2,1){15}}
\put(22,32){\vector(4,1){14.5}}
\put(22,30){\vector(1,0){14}}
\put(22,28){\vector(4,-1){14.5}}
\put(21,27){\vector(2,-1){15}}
\put(50,3){\line(0,1){8}}
\put(50,19){\line(0,1){16}}
\put(50,43){\line(0,1){8}}
\multiput(49,11)(0,8){2}{\line(1,0){2}}
\put(52,21){$\nrlv$}
\multiput(49,35)(0,8){2}{\line(1,0){2}}
\put(52,44){$\rlv$}
\multiput(48.8,10)(0.1,0){4}{\line(0,1){10}}
\multiput(57,36)(0,3){3}{\vector(1,0){14}}
\put(77,24){\framebox(27,27){Check $\ppx$}}
\end{picture}
\]
and from this experiment we get the value of $P(\ppx\vert\nrlv)$:
\[
\begin{picture}(180,55)
\put(0,30){\oval(40,40)[r]}
\put(21,33){\vector(2,1){15}}
\put(22,32){\vector(4,1){14.5}}
\put(22,30){\vector(1,0){14}}
\put(22,28){\vector(4,-1){14.5}}
\put(21,27){\vector(2,-1){15}}
\put(50,3){\line(0,1){8}}
\put(50,19){\line(0,1){16}}
\put(50,43){\line(0,1){8}}
\multiput(49,11)(0,8){2}{\line(1,0){2}}
\put(52,21){$\nrlv$}
\multiput(49,35)(0,8){2}{\line(1,0){2}}
\put(52,44){$\rlv$}
\multiput(48.8,34)(0.1,0){4}{\line(0,1){10}}
\multiput(57,12)(0,3){3}{\vector(1,0){14}}
\put(77,3){\framebox(27,27){Check $\ppx$}}
\end{picture}
\]
When we are in the classical realm, there is no need to calculate $P(\ppx)$ due to our Boolean belief revision (that is, the law of total probability):
\begin{equation}\label{eltp}
P(\ppx)=P(\ppx\vert\rlv)\,P(\rlv)+P(\ppx\vert\nrlv)\,P(\nrlv)
\end{equation}
But just for fun we may attempt to measure $P(\ppx)$ directly, removing the relevance check:
\[
\begin{picture}(180,55)
\put(0,30){\oval(40,40)[r]}
\put(21,33){\vector(2,1){15}}
\put(22,32){\vector(4,1){14.5}}
\put(22,30){\vector(1,0){14}}
\put(22,28){\vector(4,-1){14.5}}
\put(21,27){\vector(2,-1){15}}
\put(50,3){\line(0,1){8}}
\put(50,19){\line(0,1){16}}
\put(50,43){\line(0,1){8}}
\multiput(49,11)(0,8){2}{\line(1,0){2}}
\put(52,21){$\nrlv$}
\multiput(49,35)(0,8){2}{\line(1,0){2}}
\put(52,44){$\rlv$}
\multiput(57,36)(0,3){3}{\vector(1,0){14}}
\multiput(57,12)(0,3){3}{\vector(1,0){14}}
\put(77,11){\framebox(33,34){Check $\ppx$}}
\end{picture}
\]
and surprisingly discover that the result may drastically differ from \eqref{eltp}. Let us pass to exact numerical results. In order to evaluate the discrepancy, Accardi statistical invariant is used:
\begin{equation}\label{eaccdef}
\asi=
\frac{P(\ppx)-P(\ppx\vert\nrlv)}{P(\ppx\vert\rlv)-P(\ppx\vert\nrlv)}
\end{equation}
When the IR environment is classical, \eqref{eltp} holds, therefore
\[
\asi=
\frac{P(\ppx\vert\rlv)P(\rlv)+P(\ppx\vert\nrlv)P(\nrlv)-P(\ppx\vert\nrlv)}{P(\ppx\vert\rlv)-P(\ppx\vert\nrlv)}=
\]
\[
=
\frac{P(\ppx\vert\rlv)P(\rlv)-P(\ppx\vert\nrlv)P(\rlv)}{P(\ppx\vert\rlv)-P(\ppx\vert\nrlv)}=P(\rlv)
\]
that is why
\[
0\leq \asi \leq 1
\]
in classical realm. In quantum setting this is violated, see Section \ref{squant} for numerical results.

\section{Classical model: Bayesian belief revision}

In this case we suppose that the documents are like balls in an urn. 
We evaluate the probability of relevance (non-relevance, respectively) as the following ratios, introducing the notation:
\[
P(\rlv)=\frac{N_{\rlv}}{N}=\prlv\,;\quad 
P(\nrlv)=\frac{N_{\nrlv}}{N}=1-\prlv
\]
The next step of our toy scenario is to test \emph{afterwards} the occurrence of a term $X$. This gives rise to conditional probabilities, which are evaluated as follows together with the notations:
\[
P(\ppx\vert\rlv)=\frac{N_{\ppx\rlv}}{N_{\rlv}}=\qrlv\,;\quad P(\ppx\vert\nrlv)=\frac{N_{\ppx\nrlv}}{N_{\nrlv}}=\qnrlv
\]
Then apply the Bayes formula:
\[
P(\rlv\vert\ppx)=\frac{P(\ppx\vert\rlv)P(\rlv)}{P(\ppx\vert\rlv)P(\rlv)+P(\ppx\vert\nrlv)P(\nrlv)}
\]
using the parameters introduced above
\begin{equation}\label{epbx}
P_B(\rlv\vert\ppx)=
\frac{\qrlv\prlv}{\qrlv\prlv+\qnrlv(1-\prlv)}
\end{equation}
Now we are in a position to evaluate the expected boost of precision. Substituting \eqref{epbx} to \eqref{eboost29}, we have
\begin{equation}\label{eboost36}
\boost_B(\ppx)=\frac{\qrlv}{\qrlv\prlv+\qnrlv(1-\prlv)}
-1=\frac{(\qrlv+\qnrlv)(1-\prlv)}{\qrlv\prlv+\qnrlv(1-\prlv)}
\end{equation}
In the sequel, for the comparison, we shall need the expression for Accardi statistical invariant \eqref{eaccdef} for this case which for classical case is $\asi=P(\rlv)=\prlv$.

\section{Simple quantum model: spin-1/2 particle}\label{squant}

In this setting we assume that the documents form the flux of spin-1/2 quantum particles. For them, the state space is two-dimensional complex Hilbert space $\hh=C^2$. For the sake of convenience choose the properties $\rlv$ and $\nrlv$ to be basis vectors. The query state $\qss$ is a vector, denote it coordinates;
\[
\qss=\cos\frac{\phi}{2}\rlvs+\sin\frac{\phi}{2}\nrlvs
\]
Then the probabilities are expressed by the same formulas as in classical case, namely
\[
P(\rlv)=\cos^2\frac{\phi}{2}=\frac{1+\cos\phi}{2}\,;\quad
P(\nrlv)=\sin^2\frac{\phi}{2}=\frac{1-\cos\phi}{2}
\]
and the conditional probabilities for the term $\ppx$:
\[
\begin{array}{rcccccl}
P(\ppx\vert\rlv)&=&\lvert\braket{\ppx}{\rlv}\rvert^2&=&\cos^2\frac{\alpha}{2}&=&\frac{1+\cos\alpha}{2}\\
\quad\\
P(\ppx\vert\nrlv)&=&\lvert\braket{\ppx}{\nrlv}\rvert^2&=&\sin^2\frac{\alpha}{2}&=&\frac{1-\cos\alpha}{2}
\end{array}
\]
Due to the laws of quantum mechanics
\begin{equation}\label{eqtrans}
P_Q(\rlv\vert\ppx)=P(\ppx\vert\rlv)=\cos^2\frac{\alpha}{2}=\frac{1+\cos\alpha}{2}
\end{equation}
--this is because we decided to wait for the same number of documents to arrive. In `visible' terms that means that the search engine with pre-selection will work longer than without it. As a result, within our model we say nothing about recall, dealing only with precision of the IR process. The expression for the precision boost reads in this case:
\begin{equation}\label{eboostq}
\boost_Q=\frac{\cos\alpha-\cos\phi}{1+\cos\phi}
\end{equation}
Calculate the Accardi statistical invariant \eqref{eaccdef} for quantum case
\[
\asi=
\frac{\cos\phi+\cos\alpha}{\cos\alpha+\cos\alpha}
=
\frac{1}{2}\left(
1+\frac{\cos(\phi-\alpha)}{\cos\alpha}
\right)
\]
and see that it can take any real value. 

\section{Numerical simulation}

Having the two models, classical and quantum, we perform numerical simulations. We take uniformly distributed values of the parameters of both models -- classical and quantum -- and create the scatterplots, each points with coordinates $(\asi, \boost)$. The left plot corresponds to classical model (balls from an urn), and the right one depicts the results from quantum model (spin-$1/2$ particle).
\begin{tabular}{cc}
\includegraphics[width=50mm]{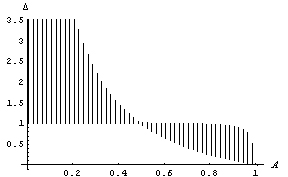}
&
\includegraphics[width=50mm]{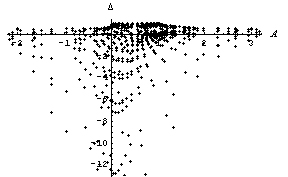}
\end{tabular}

\noindent Looking at quantum picture we see that, if the physical model is quantum, better boost is obtained by expanding queries with `non-classical' terms, those violating the Accardi restriction $0\leq \asi \leq 1$. 

Now look at the results of the experiments carried out over TIPSTER collection with the same coordinate axes \cite{melucci-x}:

\begin{center}
\includegraphics[width=50mm]{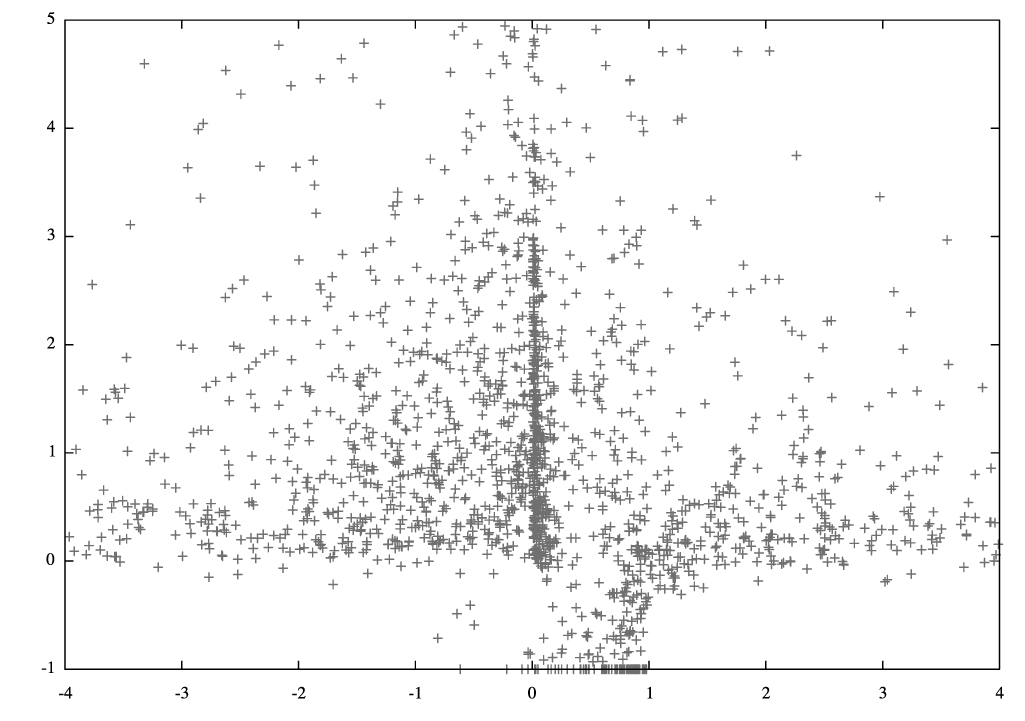}
\end{center}

These are the data based on real-world IR environment and we see that they are more similar to quantum pattern than to classical one. 

\section*{Concluding remarks}

We see that the experimental results over large data collections demonstrate features of quantum behavior. In order to mimic quantum indeterminacy, the access to complete knowledge about the system was artificially restricted. But this phenomenon is generic for Information Retrieval! The point is that search engines store some limited data about the documents rather than the documents themselves. This is a natural restriction for the access to the documents to be complete, which, in turn, could be the reason for the observed non-classicality. 

\paragraph{Acknowledgments.} The author appreciates Cris Calude, Karl Svozil and
Jozef Tkadlec for stimulating discussions on quantum contextuality during
my stay in Technical University of Vienna, supported by the Ausseninstitut
and the Institute of Theoretical Physics of the Vienna University of Tech-
nology. Many new related ideas were acquired during the Working Group Meeting `Foundations of Quantum Mechanics and Relativistic Spacetime' for COST Action MP1006, 25-26 September 2012, University of Athens, Greece.  A financial support from Russian Basic Research Foundation (grant
10-06-00178a) is appreciated.


\begin{thebibliography}{99}
\bibitem{accardi}
L. Accardi: The Axioms of Probability Theory. Conference given at the Erice school on Statistics and Probability. E. Regazzini (ed.) (1989)

\bibitem{melucci-x} M.~Melucci: An investigation of quantum interference in information retrieval.
 In {\em Proceedings of the Information Retrieval Facility Conference
  (IRFC)}, 2010


\bibitem{melucci2012}
M.~Melucci: When Index Term Probability Violates the Classical Probability Axioms Quantum Probability can be a Necessary Theory for Information Retrieval, arXiv:1203.2569 [cs.IR]

\bibitem{me-daniel}
Daniel Sonntag, Rom\`an R. Zapatrin: 
Macrodynamics of users' behavior in Information Retrieval,
arXiv:0905.2501 [cs.IR]

\bibitem{rijsbergen-book}C. van Rijsbergen: The Geometry of Information Retrieval. Cambridge University
Press, UK (2004)

\end{thebibliography}
\end{document}